\documentstyle[11pt]{article}

\newcommand{\bez}{\begin{eqnarray*}}
\newcommand{\eez}{\end{eqnarray*}}
\newcommand{\be}{\begin{equation}}
\newcommand{\ee}{\end{equation}}
\newcommand{\beq}{\begin{eqnarray}}
\newcommand{\eeq}{\end{eqnarray}}
\newcommand{\bc}{\begin{center}}
\newcommand{\ec}{\end{center}}
\newbox\grsign \setbox\grsign=\hbox{$>$} \newdimen\grdimen \grdimen=\ht\grsign
\newbox\simlessbox \newbox\simgreatbox \newbox\simpropbox
\setbox\simgreatbox=\hbox{\raise.5ex\hbox{$>$}\llap
     {\lower.5ex\hbox{$\sim$}}}\ht1=\grdimen\dp1=0pt
\setbox\simlessbox=\hbox{\raise.5ex\hbox{$<$}\llap
     {\lower.5ex\hbox{$\sim$}}}\ht2=\grdimen\dp2=0pt
\setbox\simpropbox=\hbox{\raise.5ex\hbox{$\propto$}\llap
     {\lower.5ex\hbox{$\sim$}}}\ht2=\grdimen\dp2=0pt

\def\simlt{\mathrel{\copy\simlessbox}}

\title{\vspace{4cm}\large\bf
Energy composition of the Universe:\\
time-independent internal symmetry }
\author{Arthur D.~Chernin\\
Sternberg Astronomical Institute, Moscow University, Moscow,
119899, Russia,\\
and Tuorla Observatory, University of Turku, Piikki\"o, FIN-21500,
Finland \\
}

\date{~}

\begin{document}

\maketitle

\begin{abstract}
\noindent The energy composition of the Universe, as emerged from
the Type Ia supernova observations and the WMAP data, looks
preposterously complex, -- but only at the first glance. In fact,
its structure proves to be simple and regular. An analysis in
terms of the Friedmann integral enables to recognize a remarkably
simple time-independent covariant robust recipe of the cosmic mix:
the numerical values of the Friedmann integral for vacuum, dark
matter, baryons and radiation are approximately identical. The
identity may be treated as a symmetry relation that unifies
cosmic energies into a regular set, a quartet, with the Friedmann
integral as its common genuine time-independent physical
parameter. Such cosmic internal (non-geometrical) symmetry exists
whenever cosmic energies themselves exist in nature. It is most
natural for a finite Universe suggested by the WMAP data. A link
to fundamental theory may be found under the assumption about a
special significance of the electroweak energy scale in both
particle physics and cosmology. A freeze-out model developed on
this basis demonstrates that the physical nature of new symmetry
might be due to the interplay between electroweak physics and
gravity at the cosmic age of a few picoseconds. The big
`hierarchy number' of particle physics represents the interplay
in the model. This number quantifies the Friedmann integral and
gives also a measure to some other basic cosmological figures and
phenomena associated with new symmetry. In this way, cosmic
internal symmetry provides a common ground for better
understanding of old and recent problems that otherwise seem
unrelated; the coincidence of the observed cosmic densities, the
flatness of the co-moving space, the initial perturbations and
their amplitude, the cosmic entropy are among them.



\end{abstract}
\section{Introduction}

It is increasingly evident that the Universe is fairly simple in
its overall differential geometry. Indeed, all the bulk of
observational data indicates that the co-moving space of the
Universe is uniform, isotropic and flat. To be exact, it looks
very nearly uniform, isotropic and flat. The cosmological
solutions have the simplest form for the case of a perfectly
symmetrical flat co-moving space, and these solutions provide a
good approximation to the real spacetime of the Universe.

Contrary to this, the energy content of the Universe seems to be
preposterously complex, if not absurd. According to the current
data on distant Ia type supernovae (Riess et al. 1998, Perlmutter
et al. 1999) and cosmic microwave background anisotropy (Bennett
et al. 2003, Spergel et al. 2003), the major cosmic energy
component is dark: cosmic vacuum (which is also called dark
energy) and dark matter comprise together more than 95\% (in
round numbers) the total cosmic energy. The dark sector
ingredients are well measured, but poorly understood. Indeed, the
microscopic properties of dark energy and dark matter are
unconstrained by cosmological observations and remain quite
uncertain. 'Ordinary' matter of stars in galaxies and
intergalactic gas contributes less than 5\%, and its physical
origin is almost as unclear as the nature of the dark sector. It
is embarrassing that only cosmic microwave background radiation
which contributes about 0.005 \% to cosmic energy is well
interpreted in its nature and origin.

In this paper, I review first the key observational figures that
constitute the current concordance dataset based mostly on the
WMAP observations (Sec.2). I describe also a new impressive idea
of a finite Universe derived by Luminet et al. (2003) from the
WMAP data. Then I show that a simple regularity can be recognized
behind the list of the cosmic energy ingredients. The new
regularity is found in terms of the Friedmann integral which
serves as a genuine time-independent characteristic of each of
the cosmic energies. The numerical values of the integral
estimated with the concordance data for cosmic vacuum, dark
matter, baryons and radiation prove to be approximately
identical, on the order of magnitude. The new regularity does not
depend on time, and it is valid whenever cosmic energies exist in
nature. It is a time-independent covariant recipe of the cosmic
mix (Sec.3).

This result suggests that there exists a special correspondence
among the cosmic energy ingredients which may be treated as
internal (non-geometrical) symmetry of the cosmic mix (Sec.4). New
symmetry unifies the four energy ingredients into a regular set
with a common genuine physical parameter which is the Friedmann
integral. The physical nature of cosmic internal symmetry can be
clarified under the assumption that the dark sector is well
described by `simple physics'. The simple physics assumption
adopts that dark energy is cosmic vacuum with the perfectly
uniform density $\rho_V$ and pressure $p_V$ which are also
constant in time. Cosmic vacuum with the equation of state $p_V =
- \rho_V c^2$ is equivalent to Einstein's cosmological constant,
as it was first recognized by Gliner (1965).

It is also adopted that dark matter is an ensemble of weakly
interacting massive particles (WIMPs) which have not yet
registered in laboratory experiments. It is proposed that the
WIMP mass is near 1 TeV which is the characteristic electroweak
energy scale. Accordingly, a special significance in the
framework of simple physics is prescribed to the
electroweak-scale physics. I demonstrate that the interplay
between gravity and electroweak-scale physics might be
responsible for the origin of cosmic internal symmetry (Sec.5). A
relation between new symmetry and the concept of macroscopic
extra dimensions is discussed (Sec.6). Finally, I show that
cosmic internal symmetry can shed a light on some basic features
of the real Universe that otherwise seem obscure and unrelated
(Sec.7). The results are summarized in Sec.8.

\section{The concordance figures}

The observed part of the Universe referred to as Metagalaxy
extends almost up to the principal observation horizon. The most
remote observed objects which are quasars and first galaxies are
seen at distances of ten billion light years, or $\sim 10^{28}$
cm, on the order of magnitude. The present-day horizon radius is
of the same order of magnitude:
\begin{equation}
R_0 = c t_0  = 1.2 \times 10^{28} \; cm.
\end{equation}
\noindent Here
\begin{equation}
t_0 = 13.7 \pm 0.2 \; Gyr
\end{equation}
\noindent is the current age of the Universe measured in the
proper time, according to the precision data from the WMAP
(Wilkinson Microwave Anisotropy Probe) observations. The WMAP
data provide a set of cosmological parameters measured or
constrained in observations of the temperature variations in the
cosmic microwave background (CMB). Together with the data on
cosmological supernovae  and the results of other modern
cosmological observations, the WMAP data constitute the
concordance dataset which is the empirical basis of the current
standard cosmological model.

The time rate of the evolution of the Universe as a whole, or the
rate of the cosmological expansion, is given by the Hubble
constant $H = \dot R/R$, where $R(t)$ is the cosmological scale
factor. According to the WMAP measurements in combination with
the Hubble Space Telescope (HST) observations and other studies,
the present-day Hubble constant
\begin{equation}
H_0 = 71 \pm 4 \; km/s/Mpc.
\end{equation}

With this Hubble constant, the present-day Hubble radius
\begin{equation}
R_H(t_0) = c /H_0  = 1.3 \times 10^{28} \; cm.
\end{equation}

Note a remarkable, almost exact, coincidence of the two times,
\begin{equation}
t_0 \simeq 1/H_0,
\end{equation}
\noindent and the two lengths,
\begin{equation}
R_0(t_0)  \simeq R_H(t_0).
\end{equation}

This is an example of a number of  cosmic coincidences -- some
understandable, some entirely mysterious  --  in modern cosmology.

Alongside with the cosmic age and the Hubble constant, the third
major cosmological parameter characterizing the present-day epoch
of the cosmic evolution is the dimensionless density parameter
$\Omega \equiv \rho/\rho_c$, where $\rho$ is the total density of
all the cosmic energies and $\rho_c \equiv \frac{3}{4 \pi G} H^2$
is the critical density. The value of the present-day density
parameter given by the WMAP dataset is
\begin{equation}
\Omega_0 = 1.02 \pm 0.02.
\end{equation}
\noindent According to the Friedmann theory, if the co-moving 3D
space is flat, the parameter is equal to 1. The fact that
$\Omega_0$ is exactly 1 or practically 1 today is considered a
major piece of observational evidence for the flatness of the
observed Universe. In accordance with this, the present-day
spacetime of the Universe can be described -- with good accuracy
-- by the simplest form of the metric interval:
\begin{equation}
ds^2 = c^2dt^2 - R(t)(dx^2 + dy^2 + dz^2),
\end{equation}
\noindent where synchronous proper time $t$ and the Cartesian
spatial coordinates are used. The only difference here from the
Minkowski spacetime of the Special Relativity is in the time
dependent scale factor $R(t)$ which describes the cosmological
expansion. It is the scale factor that makes the 4D spacetime
non-trivial with non-zero 4D curvature.

The present-day energy content of the Universe is characterized
by four major components. The concordance dataset provides the
current densities of cosmic vacuum (V), dark (D) matter, baryons
(B) and radiation (R):
\begin{equation}
\Omega_V = 0.66 \pm 0.07;
\end{equation}
\be \Omega_D = 0.29 \pm 0.07; \ee
\be \Omega_B h^{2} = 0.022 \pm 0.001; \ee
\begin{equation}
\Omega_R h^2 = 4.7 \alpha \times 10^{-5}, \; 1 < \alpha < 10.
\end{equation}
Here $h$ is the Hubble constant $H_0$ measured in the units 100
km/s/Mpc. Factor $\alpha$ accounts non-CMB contributions
(neutrinos, gravitons etc.) to the density of cosmic relativistic
matter.

The present-day energy densities are given by Eqs.9-12 in the
units of the present-day critical density:
\begin{equation}
\rho_c (t_0) = (3/8\pi G) H_0^2 = 0.94 \times 10^{-29} \; g/cm^3,
\end{equation}
\noindent

Note that the two major cosmic energies, vacuum and dark matter,
have comparable densities that differ in not more than a half
order of magnitude:
\begin{equation}
\Omega_V/\Omega_D = 2.6.
\end{equation}
\noindent This is one more cosmic coincidence that characterizes
the present epoch of the cosmic evolution: the time dependent
density of dark matter occurs near the time independent vacuum
density. Moreover, the two other densities, baryonic and radiation
ones, are also not too far, on the order of magnitude, from the
dominant energies. Within four orders of magnitude, all the four
energy densities are coincident at present. Why are the densities
coincident now? This is what is referred to as the `cosmic density
coincidence' problem (see Sec.8). The density coincidence is
clearly a part of a more general problem concerning the physical
nature and origin of the cosmic energies.

One of the most drastic implications from the WMAP data is the
evidence for a finite Universe (Luminet et al. 2003). The insight
into the global structure of the Universe is provided by the
power spectrum of the cosmic microwave background anisotropy at
the lowest harmonics, or the largest spatial scales. In
particular, the quadrupole is only about one-seventh as strong as
would be expected in an infinite flat space. A similar effect
(while not so dramatic) is observed also for the octopole. The
lack of power on the largest scales indicates most probably that
the space is not big enough to support them. According to Luminet
et al. (2003), the current size of the 3D co-moving volume
\begin{equation}
R_U(t_0)  = (1.03-0.82)R_0(t_0).
\end{equation}
It is very near the current horizon radius or the Hubble radius,
so that an approximate identity, $R_0 \simeq H^{-1} \simeq R_U $,
takes place at the present Universe.

A special model -- the Poincar\'e dodecahedral space of positive
spatial curvature with the density parameter $\Omega \simeq 1.013
> 1 $ -- is demonstrated to reproduce the observed shape of the
spectrum better than do models of an infinite space (Luminet et
al. 2003; Luminet 2005, Aurich et al. 2004). The figure for the
density parameter is compatible with the WMAP limitations $\Omega
= 1.02 \pm 0.02$ (see Eq.7). The idea of a finite Universe -- even
independently of the special model with its rigid geometry and
fixed parameters -- can be verified by further analysis of the
WMAP data and the upcoming Planck data. If confirmed, this is a
major discovery about the nature of the Universe (Ellis 2003).

\section{Recipe of cosmic mix}

The list of the energy ingredients (Eqs.9-12) can be rewritten in
terms of the Friedmann integral (hereafter FINT) which  is a
constant genuine physical characteristic of each of the cosmic
energies. The FINT enables one to eliminate the effect of
cosmological expansion from the description of the cosmic energy
composition. Historically, the quantity appeared in Friedmann's
first paper on cosmological expansion (Friedmann 1922) where the
length-dimension constant $A$ was introduced to represent
non-relativistic matter in the dynamical equation for the
cosmological scale factor.

The FINT comes from the Friedmann `thermodynamical' equation which
is applied to each of the energy ingredients individually:
\be \frac{\dot \rho}{\rho(1 + w)} = - 3 \frac{\dot R}{R}. \ee
\noindent Here the constant pressure-to-density ratio $w = p/\rho
= -1, 0, 0, 1/3$ for vacuum, dark matter, baryons and radiation,
respectively; $R(t)$ is the cosmological scale factor. The
integral of the equation may be given in the form
\begin{equation}
A = [\kappa \rho R^{3(1+w)}]^{\frac{1}{1+3w}},
\end{equation}
where $\kappa = \frac{8 \pi G}{3 c^2}$ and $G$ is the
gravitational constant.

Because of their origin from Eq.16 as constants of integration,
the values of the FINT for vacuum, dark matter, baryons and
radiation are completely independent of each other {\it a priori}
and not restricted by any theory constraints (except for trivial
ones).

In the Friedmann dynamical equation, the four FINT values $A_V,
A_D, A_B, A_R$ represent vacuum, dark matter, baryons and
radiation, respectively:
\begin{equation}
{\dot R/c}^2 = (A_V/R)^{-2} + A_D/R + A_B/R + (A_R/R)^{2} - K.
\end{equation}
\noindent Here $K$ is the constant which is zero in a model of a
flat 3D space. In models of a non-zero 3D curvature, the scale
factor $R(t)$ is usually identified with the curvature radius $a
(t)$, and then $K = 1, -1$, in Eq.18. In a finite-size Universe
with positive spatial curvature (as in the model by Luminet et
al. 2003), the scale factor $R(t)$ is most naturally identified
with the finite size $R_U(t)$ of the 3D space; in this case, $K
=(R/a)^{2} = Const > 0$.

It is seen from Eq.17, that the value of the FINT for vacuum does
not depend on the scale factor and its normalization; this is a
universal constant which is the same in any cosmological model
and in any reference frame:
\begin{equation}
A_V = (\kappa \rho_V)^{-1/2} \simeq \Omega_V^{-1/2} c/H \simeq 1
\times 10^{28} \; cm.
\end{equation} \noindent

We see now a new remarkable cosmic coincidence -- the universal
constant length $A_V$ turns out to be very near the present-day
values of the lengths $R_0, R_H$ and $R_U$:
\be R_0(t_0) \sim R_0(t_0) \sim R_U(t_0) \sim A_V. \ee This triple
coincidence is not too mysterious, as we will see in Sec.7.

The FINT values for non-vacuum energies depend on the scale factor
and its normalization explicitly. If the Universe is really
finite in size, we may use the most natural scale-factor
normalization to the size of the cosmic space:
\be R(t) = R_U(t) \simeq A_V (1+z)^{-1}.\ee

With this normalization, the FINT non-vacuum values have a clear
physical sense. Indeed, the values $A_D$ and $A_B$ are determined
by the total masses of dark matter, $M_D$, and the total mass of
baryons, $M_B$, respectively:
\begin{equation}
A_D = 2 \kappa M_D, \;\; A_B = 2 \kappa M_B.
\end{equation}
\noindent

The FINT value for radiation is determined by the total number of
the CMB photons (and other possible relativistic particles),
$N_R$, in the finite-size Universe:
\begin{equation}
A_R \simeq (\kappa \hbar c)^{1/2} N_{R}^{2/3},
\end{equation}
\noindent where $\hbar = h/2\pi$, and $h$ is the Planck constant.

The normalization of Eq.21 may be used as well, if the co-moving
space is infinite; in this case, it may be considered as the
normalization to the size of the Metagalaxy. Then the constant
total figures $M_D, M_D$ and $N_R$ will be related to the whole
visible space. The quantitative results for the FINT (see below)
will be the same in both cases, since the size of the Metagalaxy
is near both $ct_0$ and $A_V$, at the present epoch.

With the data of Eqs.10-12 and the normalization of Eq.21, the
FINT non-vacuum values are
\begin{equation}
A_D =  \kappa \rho_D R^3 \simeq \Omega_D R^3 H^2 \simeq \Omega_D
c/H \simeq 3 \times 10^{27} \; cm.
\end{equation}

\begin{equation}
A_B = \kappa \rho_B R^3 \simeq \Omega_B R^3 H^2 \simeq \Omega_B
c/H \simeq 3 \times 10^{26} \;cm.
\end{equation}

\begin{equation}
A_R = (\kappa \rho_R )^{1/2} R^2 \simeq (\Omega_R \alpha)^{1/2}
c/H \simeq 1 \times  10^{26} cm, \;\;\; (\alpha \simeq 1).
\end{equation}

Eqs.20,24-26 give a time-independent recipe of the cosmic mix.
The recipe proves to be simple -- all the FINT values are nearly
identical, on the order of magnitude:
\begin{equation}
A_V \sim A_D \sim A_B \sim A_R  \sim 10^{27 \pm 1} cm \sim 10^{60
\pm 1} M_{Pl}^{-1}.
\end{equation}
Here the `natural units' are used in which the speed of light, the
Boltzmann constant and the Planck constant are all equal to
unity: $c = k = \hbar = 1$. The Planck mass  $M_{Pl} = G^{-1/2}
\simeq 1.2 \times 10^{19}$ GeV. Though the FINT identity of Eq.27
is found with the data on the present (special) epoch of cosmic
evolution, it is valid for all the epochs whenever the four
energies exist in nature.

The FINT identity for radiation and `ordinary matter' was first
found (Chernin 1968) soon after the CMB discovery. With the
discovery of dark matter and cosmic vacuum, the identity was
extended to all the four energies (Chernin 2001) (in these two
works -- contrary to the present one, -- a normalization of the
scale factor to the curvature radius was used).

The result of Eqs.20, 24-26 may be rewritten as a set of four
dimensionless constant quantities defined as follows:
\be mix \equiv 10 A/A_V.\ee With the figures above, we have:
\be mix \simeq [10, 3, 0.3, 0.1], \ee \noindent where the four
dimensionless numbers relate to vacuum, dark matter, baryons and
radiation, respectively.  These (somewhat `rounded up') numbers
are all of the order-of-unity, if one agrees, as usual, that a
number between 0.1 and 10 is of the unity order.

\section{Cosmic internal symmetry}

According to one of the most general definitions, any symmetry
describes a similarity of objects in a set (Weyl 1951). If
symmetry does not concern spacetime relations, it is referred to
as internal symmetry  -- in contrast to geometrical symmetries. A
typical example of internal symmetry is symmetry between the
proton and the neutron: the particles differ in mass, electric
charge, life-time, etc., but they constitute a set which is a
hadron doublet with a common constant value (1/2) of isotopic
spin.

In the same way, the FINT identity of Eq.27 describes the
similarity of the four cosmic energy ingredients. This similarity
may be referred to as `cosmic internal symmetry' (hereafter
COINS). The energy ingredients are obviously different in many
respects, and it is most essential that one of them is vacuum,
while the three others are non-vacuum energies. Despite this and
other differences, the four energies constitute a regular set --
a quartet -- with the Friedmann integral $A$ as its common
(approximately identical for all the members) genuine constant
physical parameter.

Briefly, some major features of new symmetry:

1. COINS is time-independent symmetry in the evolving Universe. It
exists at least since the earliest epoch of the cosmic history
which can be traced with the current observational data -- this is
the epoch of the Big Bang Nucleosynthesis. The FINT for baryons
exists  since the epoch of $\sim 1$ GeV temperatures, redshifts
$\sim 10^{12}$ and the cosmic age $\sim 10^{-6}$ sec when baryons
became non-relativistic. In the future, it exists until the decay
of the proton, i.e. to the cosmic age of $ \ge 10^{32}$ years. It
means that the FINT for baryons is the same over 45 decades of
the cosmic time. The FINTs for vacuum and radiation are constant
even for longer times: they are practically eternal. The FINT for
dark matter exists for all the cosmic past when the dark matter
particles are non-relativistic; it is since the cosmic age of a
few picoseconds, if the exist are `weakly interacting massive
particles' (WIMPs) with a mass near 1 TeV (see below).

2. COINS is covariant symmetry, since it is formulated in terms of
the FINT which is determined by the scalar (invariant) quantities
$M_D, M_B, N_R, \rho_V$ . The FINT is associated with the
four-dimensional Riemann invariant, ${\mathcal{R}} = 8 \pi G
(\rho - 3 p)$. In the limit of infinite time ${\mathcal{R}}
\rightarrow 32 \pi G \rho_V = 12/A_V^2, \; t \rightarrow \infty$.
If cosmic matter is initially generated in the form of massless
particles (and the particles acquire mass later via, say, the
Higgs mechanism), the invariant is the same in the opposite time
limit as well: $RI \rightarrow 12/A_V^2, \; t \rightarrow 0$.
Identified in the co-moving space, COINS exists in any other
spatial sections and in the 4D spacetime as a whole.

3. COINS is not exact, but approximate symmetry, since the four
FINT values differ within two orders of magnitude. At fundamental
level, its violation might be related to, for instance,
fundamental particle-antiparticle asymmetry which is most
probably involved in baryogenesis.

4. COINS implies that there is a correspondence between the total
dark matter mass, the total baryonic mass and the total number
$N_R$ of relativistic particles (the CMB photons):
\be N_R^{2/3} \sim M_D/M_{Pl} \sim M_B/M_{Pl}. \ee

5. COINS implies also that there is a consistency of extensive
cosmic quantities $M_D, M_B$, $N_R$ and the intensive quantity
$\rho_V$:
\be \rho_V \sim (M_{Pl}/M_D)^2 M_{Pl}^4 \sim (M_{Pl}/M_B)^2
M_{Pl}^4 \sim N_R^{-4/3} M_{Pl}^4. \ee

6. Due to COINS, the Friedmann dynamical equation (Eq.18) contains
not four empirical energy parameters, but (as it was said in the
section above) in fact only one universal empirical parameter $A$
which is the Friedmann integral common for all the four energy
ingredients -- in the first and main approximation.

\section{Gravity-electroweak interplay}

What is the physical nature of new symmetry? It is obvious that
the real understanding of the problem is hardly possible now
because the origin of cosmic vacuum, dark matter and baryons is
yet completely unknown. However there is a reasonable approach to
the problem: this is the assumption that the cosmic energy
ingredients are well described by `simple physics'. A simple
physics approach adopted here assumes that dark energy is vacuum
with constant density and $w = -1$ (as above). Also dark matter
is WIMPs which are stable or long-living thermal relics of the
early Universe. Under these (and some other -- see below)
assumptions, a model can be developed that describes how, in
principle, COINS might originate in the early Universe.

The model addresses physical processes at the epoch of
electroweak-scale temperatures, $T \sim M_{EW} \sim 1$ TeV. A
reason for that is the special significance of the electroweak
energy scale in fundamental physics (Okun 1985, Rubakov 1999,
Weinberg 2000). At the epoch of TeV temperatures, the cosmic age,
$t_{EW}$, is about a few picoseconds, and the horizon radius,
$R_{EW}$ is of a fraction of 1 mm.

Two major factors are involved in the model: electroweak-scale
physics and gravity which controls the rate of the cosmological
expansion. They are represented in the model by two fundamental
constants which are the electroweak energy/mass $M_{EW}$ and the
Planck mass $M_{Pl} = G^{-1/2}$. The model describes the COINS
origin as a result of the interplay between gravity  and
electroweak physics. The gravity-electroweak interplay reveals
itself in the WIMP freeze-out at the $M_{EW}$ temperatures.

The cosmological freeze-out kinetics is well-known (see Zeldovich
\& Novikov 1982, Dolgov et al. 1988, Kolb \& Turner 1990). In a
simple version suggested by Arkani-Hamed et al. (2000), the WIMPs
freeze out when the temperature $T$ falls to the particle mass
$m$ and the expansion rate $1/t$ wins over the annihilation rate,
$\sigma n$. Here the annihilation cross-section $\sigma \sim
m^{-2}$ and $n$ is the number density of particles. Accordingly,
at that moment,
\begin{equation}
n \sim 1/(\sigma t) \sim m^2 (G \rho_R).
\end{equation}
\noindent The approximate cosmological relation $t \sim (G
\rho_R)$ is also used for the early radiation domination epoch.

Introducing the FINT values for dark matter, $A_D$, and for
radiation, $A_R$ and putting $\rho_D \sim m n$, one finds:
\begin{equation}
A_D \sim R(t) m^3 M_{Pl}^{-2} A_R.
\end{equation}
\noindent One also has at that moment $\rho_R \sim T^4 \sim m^4$,
and because of this
\begin{equation}
 A_R \sim  R(t)^2 m^2 M_{Pl}^{-1},
\end{equation}
\noindent where $R(t) \sim A_V (1 + z)^{-1}$ is the scale factor
(normalized as in Eq.28), and $z$ is the redshift, at the
freeze-out epoch. The system of Eqs.32-34 describes the
freeze-out kinetics in terms of the FINT values $A_D, A_R, A_V$.

If the system has a solution in terms of $M_{Pl}$ and $M_{EW}$
only, the vacuum density must be (Arkani-Hamed et al. 2000)
\begin{equation}
\rho_V \sim (M_{EW}/M_{Pl})^8 M_{Pl}^4.
\end{equation}
\noindent

With this density, the vacuum integral is
\begin{equation}
A_V \sim  (M_{Pl}/M_{EW})^4 M_{Pl}^{-1}.
\end{equation}

Arguing along this line, one might expect that the mass of the
particle must be identified with $M_{EW}$ (the other mass
$M_{Pl}$ is enormously large for this) and the redshift $z$ at
the freeze-out epoch is a simple combination of the two energy
scales:
\begin{equation}
z \sim M_{Pl}/M_{EW}.
\end{equation}

Then one has the solution of the system:
\begin{equation}
A_M \sim A_R \sim A_V \sim  (M_{Pl}/M_{EW})^4 M_{Pl}^{-1}.
\end{equation}

Thus, the equality of the three FINT values appears as an outcome
of the interplay between gravity and electroweak-scale physics
which controls the freeze-out kinetics. The model gives also the
Friedmann integral in terms of the two fundamental energy scales
$M_{Pl}$ and $M_{EW}$.

To refine the quantitative estimates, one may introduce, as usual,
a `reduced Planck scale' $\bar M_{Pl} = 0.1 M_{Pl}$ which takes
into account the effective number of the degrees of freedom that
must be included in the freeze-out kinetics and also factors like
$8 \pi/3$ or $32 \pi/3$ in exact cosmological formulas (see
Zeldovich \& Novikov 1982, Kolb \& Turner 1990).  Then one gets:
\begin{equation}
 A \sim (\bar M_{Pl}/M_{EW})^4 M_{Pl}^{-1} \sim 10^{60} M_{Pl}^{-1}.
\end{equation}
\noindent A quantitative agreement with the empirical result of
Eq.27 is quite satisfactory here.

The big dimensionless ratio
\be X = \bar M_{Pl}/M_{EW} \sim 10^{15} \ee \noindent that enters
the result is known as the  hierarchy number, in particle
physics. It characterizes the huge gap between the two
fundamental energies. The nature of the gap is not well
understood, and this is considered as one of the most difficult
problems in fundamental theory (see more about this in Sec.5
below).

Note that the freeze-out model is not complete: it does not
account for the FINT value for baryons. It may, however, be
assumed, that baryons can be included in a more general model of
gravity-electroweak interplay, that assumes that baryogenesis
takes place at the electroweak temperatures. Electroweak
baryogenesis was proposed by Kuzmin et al. (1985); see also
Dolgov (1992) and Rubakov (1999) for a critical review of the
problem.

Moreover, the gravity-electroweak interplay might also be
responsible for the origin of cosmic vacuum via supersymmetry
violation at TeV temperatures (see Zeldovich 1968, Dolgov 2004 and
references therein). If so, the epoch of electroweak energies is
the real beginning of the evolution described by the current
standard cosmological model. At that epoch, cosmic energies come
into existence due to a common physical process. Their common
origin in the `Electroweak Big Bang' might guarantee, in
particular, the internal mutual correspondence among them which
manifests itself as COINS, at phenomenological level.

\section{COINS and extra dimensions}

As is seen from the results of the section above, the hierarchy
number $X$ provides COINS (and associated phenomena -- see the
next section) with a common quantitative measure. It may be
expected that if the hierarchy problem is resolved in fundamental
physics, it gives a new insight into the nature of COINS and the
cosmic energy origin.

Presently, the hierarchy problem is completely open. There is
however an interesting recent approach to its understanding that
may be useful for cosmology. Arkani-Hamed et al. (1998) proposed
an idea of macroscopic extra dimensions to eliminate the energy
hierarchy of fundamental theory. The idea assumes that there
exist finite (compactified) macroscopic spatial extra dimensions
in space. In a finite Universe, the extra dimensions constitute,
together with the 3D space, a close multi-dimensional space. This
multi-dimensional space is treated as the `true space of nature'.

It is also assumed that there is one and only one `truly
fundamental' energy scale $M_*$ in nature, and  that this scale is
close to the electroweak scale $M_{EW}$. As for the Planck scale,
it is reduced to a combination of the scale $M_*$ and the size
$R_*$ of the compact macroscopic extra dimensions of the true
space:
\begin{equation}
M_{Pl} \sim (M_{*} R_*)^{n/2} M_{*}.
\end{equation}
\noindent Here $n$ is the number of the extra dimensions, which
are proposed to be of the same size. Together with the Planck
mass, the gravitational constant in three-dimensional space, $G =
M_{Pl}^{-2}$, looses its fundamentality and is reduced to a
combination of the two truly fundamental constants $M_*$ and
$R_*$.

It is reasonably argued  that the case $n = 2$ is the most
appropriate one; if so, the size of two extra dimensions is in
the millimeter (or submillimeter) range:
\begin{equation}
R_* \sim 0.1 \;\;cm, \;\;\; n = 2.
\end{equation}

It is clear that when the hierarchy number, $ X = M_{Pl}/M_{EW}$,
is replaced with the product
\be X = M_* R_*, \ee

this is not elimination of the hierarchy, but its re-formulation
in the new terms of $M_*$ and $R_*$.

In the multi-dimensional space, all the physical fields, except
gravity, are assumed to be confined in the three-dimensional
space, or brane. The multi-dimensional physics affects the brane
via gravity, and therefore cosmology must be re-formulated in
terms of the true fundamental constants. In particular, one has
from Eq.38 for the FINT:
\begin{equation}
A \sim (M_{*} R_*)^{(3/2)n} M_{*}^{-1}.
\end{equation}
\noindent In the case of two extra dimensions:
\begin{equation}
A \sim (M_{*} R_*)^{2} R_*, \;\;\;\; n = 2.
\end{equation}

Then the vacuum density
\begin{equation}
\rho_V  \sim (M_{*} R_* )^{-2n} M_{*}^{4},
\end{equation}
and in the case of two extra dimensions:
\begin{equation}
\rho_V \sim R_*^{-4}; \;\; n = 2.
\end{equation}

This is a surprising result: the vacuum density proves to be
expressed via the size of the extra dimensions alone. The new
relation is free from any signs of the hierarchy effect (Chernin
2002b). In this important case, the hierarchy is really eliminated
from the multi-dimensional physics.

According to the idea of extra dimensions, all we observe in
three-dimensional space are shadows of the true multi-dimensional
entities. In particular, it may be assumed that true vacuum
exists in the multi-dimensional space, and the observed cosmic
vacuum is not more than its 3D projection to the cosmological
brane. This is possible, only if vacuum is due to gravity alone
and not related to the fields of matter. In this case, the true
vacuum is defined in the multi-dimensional space, and its density
$\rho_{V5} \sim R_*^{-6}$, for two extra dimensions. This `true
vacuum' is also free from the hierarchy effect.

But if the observed vacuum density is due to supersymmetry
violation (Zeldovich 1968), vacuum is confined in the brane --
together with fermionic and bosonic fields of matter. In such a
case, a real sense of the relation between the vacuum density and
the extra-dimension size would be not obvious.

Thus, taking the relations of this section at face, one may
conclude that the basic cosmological parameter FINT and also
$\rho_V$  have roots in the extra-dimension physics, -- if extra
dimensions really exist.

It is expected that the idea of macroscopic extra-dimensions will
be directly tested with the Big Hadron Collider and in
submillimeter laboratory experiments in the coming several years
-- perhaps, at the same time when the Planck mission will test
the compactness of the finite 3D space on cosmological scales.

\section{COINS related figures and phenomena}

Cosmic internal symmetry offers a productive common ground for
better understanding of a number of cosmological problems that
otherwise seem unrelated. The problems concern a wide range of
basic figures and phenomena.

\subsection{Cosmic density coincidence}

According to the WMAP results and all the concordance data
(Sec.2), the densities of the two energies in the dark sector of
the cosmic mix, $\rho_V$ and $\rho_D$, are nearly coincident at
the present epoch. Why should we observe them to be so nearly
equivalent right now? While the vacuum density (or the
cosmological constant) is by definition time independent, the
dark matter density is diluted as $R^{-3}$ as the Universe
expands. Despite the evolution of $R(t)$ over many orders of
magnitude, we appear to live at an epoch during which the two
energy densities are roughly the same. This is the `cosmic
coincidence' problem which is commonly considered as a severe
challenge to the current cosmological concepts (see, for
instance, Chernin, 2002, and references therein).

Note that the idea of quintessence was initially introduced in an
attempt to eliminate the problem. However it is now clear that
quintessence can hardly be useful because the pressure-to-density
ratio has recently been found to lie between -1.2 and -0.9
(Perlmutter et al. 2003), which seemingly rules out the idea.
Contrary to this, COINS suggests a natural solution to the
problem without any additional assumptions.

In a broader view, all the four energy densities, the two dark
ones and the two others, are of nearly the same order of
magnitude. Their coincidence is temporary and therefore
accidental, in this sense. The `eternal' coincidence of the FINT
values is really behind it.

Indeed, taking the approximate identity of the Friedmann
integrals as a basic relation, one has for the four densities:
\begin{equation}
\rho_V \sim (M_{Pl}/A)^{2}, \; \; \rho_D \sim A/R^3 M_{Pl}^{2},
\;\; \rho_B \sim A/R^3 M_{Pl}^{2}, \; \; \rho_R \sim A^2/R^4
M_{Pl}^{2}.
\end{equation}
It is seen from these equations that the four densities must
become identical (approximately) and equal to $\sim (M_{Pl}/A)^2$,
when $R(t_0) \sim A$.

Thus, the four cosmic densities are near coincident because of
COINS and the special character of the moment of observation at
which the size of the finite Universe and/or the size of the
Metagalaxy are equal to the Friedmann integral.

In fact, all the coincidences that take place at the present
epoch (Sec.2) are due to the only equality $R(t_0) \sim A$.  It
follows from this equality that $\rho_V \sim \rho_D(t_0)$ (see
Eq.48); but this means that the present epoch is the epoch of
transition from the matter domination to the vacuum domination.
At this epoch, the solution for the matter domination, $R(t)
\propto t^{2/3}$, and the solution for the vacuum domination,
$R(t) \propto \exp(ct/A_V)$, are both valid, in a rough
approximation. We have from the first and second of them,
correspondingly:
\be H (t_0) \simeq 2/(3t_0); H (t_0) \simeq c/A_V. \ee Then the
equality $t_0 \sim H(t_0)^{-1}$ comes from these two relations
directly. For a finite space, we have also from this that $R_U
\sim ct$ at present.

The fact that the size of the finite space is near the Hubble
radius at present is sometimes treated as a strange accident or
unnatural tuning in the finite-space model by Luminet et al.
(2003). As we see now, this is not the case. Actually when the
size of the space reaches the universal constant $A_V$, the size
turns to be necessarily and naturally near the Hubble radius. This
consideration eliminates a critical argument against the model.

Another question is why we happen to live at such a special
epoch. This is among the matters that are effectively discussed
on the basis of the Anthropic Principle (see, for instance,
Weinberg 1987).

\subsection{The Dicke problem}

The geometry of the co-moving space looks nearly flat in
observations, and the cosmological expansion proceeds in a nearly
parabolic regime. The both is quantified by the density parameter
$\Omega (t)$ which is measured to be near unity (see Sec.2). Why
this is so? The question is known as the `flatness problem' that
was first recognized by Dicke (1970) who mentioned that the
Universe must be extremely finely tuned to yield the observed
balance between the total energy density of the Universe and the
critical density.

In the 1970-s, the observational constraints on $\Omega (t_0)$
were much weaker than now, and it was considered that this
quantity was between 0.1 and 10. Such an apparently wide range
implies a very narrow range at earlier epochs. It was estimated
that the density balance quantified by $\Omega$ must be tuned
with the accuracy $\sim 10^{-16}$ or $\sim 10^{-60}$, if it is
fixed at the epoch of the Big Bang Nucleosynthesis (BBN) or at
the Planck epoch, respectively. Such a fine tuning in the `initial
conditions' for the cosmological expansion was reasonably
considered by Dicke as unacceptable (see, for instance, Chernin
2003 for more references).

COINS shows the Dicke problem in quite different light. Indeed,
the correspondence between vacuum and dark matter described by
the symmetry relation $A_V \sim A_D$ puts a strong upper limit to
any deviations from the flatness in possible models with non-zero
spatial curvature. The deviations are measured by the quantity
$\vert \Omega (t) -1\vert$, and, as is seen from the Friedmann
equation of Sec.3, this quantity goes to zero in both limits $t
\rightarrow 0$ and $t \rightarrow \infty$. At earlier epoch, the
deviations are restricted by the matter gravity, and at the later
epoch, they are restricted by the vacuum antigravity. The extreme
deviation takes place in the era when gravity and antigravity
balance each other. The corresponding redshift
\be 1 + z  = 1 + z_V \simeq (2 A_V/A_D)^{1/3} \simeq 1. \ee At
that time,
\begin{equation}
\Omega (z_V) -1  \simeq [1 \pm \frac{1}{2} (
\frac{A_V}{A_D})^{2/3} (R/a)^{2}]^{-1} -1 \simeq \pm \frac{1}{2}
(A_V/A_D)^{2/3} (A_V/a_0)^2.
\end{equation}
\noindent Here $R(t)$ is the scale factor normalized as $R(t) =
A_V(1 + z)^{-1}$ (see Eq.21), and $a_0$ is the present-day space
curvature radius.

As we see, there is the upper limit for any possible nonflatness
at the present, in the past and future of the Universe:
\be \vert \Omega (z) -1 \vert \le \vert \Omega (z_V) -1 \vert
\simeq \frac{1}{2}(A_V/A_D)^{2/3} (A_V/a_0)^2. \ee

Nonflatness is quantified by the constant parameter $y \equiv
\frac{1}{2}(A_V/A_D)^{2/3} (A_V/a_0)^2 \simeq (A_V/a_0)^2$ which
might be fixed by initial conditions at the TeV temperature
epoch, at the BBN epoch, at the Planck epoch or at any other
epoch equally, because the parameter is time-independent. The
parameter is also normalization-independent.

Any cosmological model of non-zero spatial curvature fits the
1970's observational constraints, if the parameter $y = \simlt
1$. The modern WMAP constraints are met, if the parameter $y
\simlt 0.02$. For the Luminet's et al. (2003) model with $\Omega
= 1.013$ we have $y \simeq 0.1$. To see the contrast with the
fine-tuning argument, one may compare modest numbers like 1 or
0.02-0.01 with the enormous numbers $10^{-16}$ and $10^{-60}$. A
similar result has recently been found in a complementary
treatment by Adler and Overduin (2005).

Thus, the balance between vacuum antigravity and dark matter
gravity is actually behind the observed near flatness of the 3D
co-moving space. This balance is controlled by COINS which rules
out any significant deviations from flatness at any time.

Note that no special hypothesis (about, say, an enormous vacuum
density, or enormous energy density of inflanton field, at
enormously large $z$) is required to clarify and eliminate the
Dicke fine-tuning problem. The really observed vacuum density and
the standard cosmology at modest $z$ are quite enough to
understand why the observed space is nearly flat and the cosmic
expansion is nearly parabolic.

\subsection{Perturbation amplitude}

A fine-tuning problem which is similar to the Dicke argument is
well-known in the theory of cosmic structure formation. Indeed,
the perturbations must be extremely finely tuned in amplitude to
come to the nonlinear regime between the red shifts, say, $z
\simeq 3-10$ (when the oldest galaxies are observed) and $z = z_V
\simeq 1$ (when the vacuum antigravity terminates the linear
perturbation growth -- see, for instance, Chernin et al. 2003).
Consider, for example, the large-scale adiabatic perturbations
which are ever grow before $z \sim 1$. Using the standard theory
of weak perturbations (Lifshits 1946), we may easily see that,
the perturbation generated at the BBN epoch must increase
$10^{16}-10^{17}$ times, so that their initial amplitudes must be
tuned with the accuracy better than $10^{-16}$ to guarantee
nonlinearity in the appropriate redshift range. If perturbations
are generated at the Planck epoch, the accuracy must be better
than $10^{-60}$.

The numerical similarity with the Dicke considerations is not
purely accidental. It is long known due to Zeldovich (1965) that
the correct time rate of the perturbation growth could be
obtained in a simple picture in which a perturbation overdensity
is treated as a part of a universe of positive curvature on the
unperturbed background of a flat space. The relative amplitude of
density perturbation is given in this case by the deviation of the
density parameter $\Omega$ from unity:
\begin{equation}
\delta \equiv \delta \rho/\rho \simeq \Omega (t) -1.
\end{equation}
\noindent It is because of this relation that the perturbation
growth resembles the evolution of nonflatness. We will show now
that this analogy may help to understand the nature of the initial
perturbation amplitude which is a key quantity in the theory of
structure formation.

Following Zeldovich (1965), we may generally assume that various
perturbation areas are described by models with different
curvature parameters $K >0, K = 0, K < 0$ (see Sec.3) and
different curvature radii $a(t)$, but with the same set of the
Friedmann integrals as in the background model. If, for instance,
$K = 0$ in the background model, then areas with $K > 0$ and $K <
0$ correspond to over-density perturbations and under-density
perturbations, respectively. The perturbation areas of various
sizes $r(t)$ develop independently of each other (even if they
spatially overlap), in the linear approximation.

In such a simple example, we will use the parabolic ($K = 0$)
solution with the scale-factor normalized as $R(t) \simeq A_V (1 +
z)^{-1}$ to describe the unperturbed background expansion. Then
an overdensity perturbation may be described by a model with $K >
0$ normalized in the same manner. In accordance with Eq.53 and the
results of the subsection above, the density contrast $\delta$ in
dark matter reaches its maximum when the redshift $z = z_V \sim
1$. At that time
\begin{equation}
\delta (z_V) \equiv  \vert \delta \rho_D/\rho_D \vert  = \vert
\Omega_{z_V} - 1 \vert \simeq \vert [1 \pm \frac{1}{2}(
\frac{A_V}{A_D})^{2/3} (A_V/a_0)^{2}]^{-1} -1\vert.
\end{equation}
\noindent Here $a_0$ is the present-day value of the curvature
radius corresponding to a given perturbation area.

The value of the amplitude $\delta(z_V)$ is about unity, $\delta
\sim 1$, provided the constant parameter $y =
\frac{1}{2}(A_V/A_D)^{2/3} (A_V/a_0)^2 \sim 1$.

No fine tuning of the amplitude is needed, as we see: the
order-of-unity constant parameter $y$ guarantees the
quantitatively correct perturbation evolution. This parameter may
be fixed by the `initial conditions' at any epoch in the past,
because the parameter is time independent.

Remind that the result relates to the perturbations which grow
all the time in the past when $z \ge z_V$. The spatial scales of
these perturbations are large enough: they are ever not less than
the Jeans critical length for gravitational instability $R_J(t)$.
The Jeans length has maximum at the moment $z = z_*$ when the
matter density $\rho_D + \rho_B$ is equal to the radiation density
$\rho_R$ (see, for instance, Zeldovich and Novikov 1983). At this
moment, $z_* \simeq A_V A_D/A_R^2$, and
\be R_J(t_*) \simeq ct_* \simeq 0.4 A_R^3/A_D^2. \ee

The minimal spatial scale $L(t_*)$ for ever growing perturbations
is near the Jeans length, at this moment: $L(t_*) \simeq
R_J(t_*)$. At $z = z_V$, this scale is $L (z_V) = R_L(t_*)(1 +
z_*)/(1 + z_V) \simeq 0.2 A_V A_R/A_D. $

With this relation, we may estimate the amplitude of the
perturbation of the gravitational potential $\Delta$, at this
scale. In accordance with the general theory (Lifshits 1946), we
find:
\be \Delta = \delta (z_V)[L(z_V)/ct_V]^2 \simeq 0.2 \delta (z_V)
(A_R/A_D)^2, \ee \noindent where $t_V = t(z_V) \simeq 0.5 A_V/c$.

Since $\delta (R_L, z_V) \simeq 1$, the value $\Delta$ turns out
to be expressed in terms of the Friedmann integrals only:
\be  \Delta \simeq 0.2 (A_R/A_D)^2 \simeq 10^{-4}.\ee

The general theory indicates that this value does not depend on
time: $\Delta = Const(t)$. Moreover, this value is scale
independent initially, if the initial perturbations have the
Harrison-Zeldovich (HZ) spectrum: $\delta \propto r^{-2}.$ The HZ
spectrum is in good agreement with the WMAP data (Spergel et
al.2003).

Thus, the key quantity of gravitational instability comes in a
quite natural and simple way as a universal dimensionless
constant of cosmology $\Delta$. Together with the HZ spectrum,
this constant gives a complete quantitative description of the
initial adiabatic perturbations that seed the large-scale cosmic
structure.

No fine tuning for the amplitude is needed at all. Since the
quantity $\Delta$ is a constant, the initial conditions for the
perturbations do not need to be fixed at any specific `initial'
moment: we have in fact `no-initial conditions' situation.

In the context of the gravity-electroweak interplay (Sec.5), it is
especially interesting to estimate the perturbation amplitude at
the epoch when the cosmic temperature $\sim M_{EW}$ and the red
shift $z = z_{EW} \sim X = \bar M_{Pl}/M_{EW}$. With the
relations above, we find for the scale $r = L$:
\begin{equation}
\delta (z_{EW}, L) \simeq \Omega (z_{EW}) -1 \simeq
[R(t_0)/A_R]^2(1 + z)^{-2} \sim X^{-2} \sim 10^{-30}.
\end{equation}

As we see, the perturbation amplitude at $z = z_{EW}$ is given in
terms of the universal hierarchy number $X$ alone. In this way,
COINS together with the freeze-out physics provide the
perturbations with a natural initial amplitude.

We can refine somewhat the quantitative results with the use of
the concordance data. It has been long known and confirmed
recently by the WMAP data (Spergel ett al. 2003), that the largest
scale at which the density perturbations have the unity amplitude
is $r_1 \simeq 8 h^{-1}$ Mpc, at present. At $z= z_V$, this scale
$r_1(z_V) \simeq 4$ Mpc. This is smaller than $L(z_V)$, which
means that the scale $r_1 (t)$ is in the range where the HZ
spectrum leads to an almost scale-independent amplitude at $z =
z_*$:
\be \delta (r) \propto (1 + 2 \ln R_L/r), \;\;\; r < L, \;\;\;
z=z_* . \ee

This spectrum keeps the same shape to the moment $z = z_V$ at
which
\be \delta (r_1) = \delta (R_L) (1 + 2 \ln R_L/R_1) = 1, \;\;\;
z=z_V. \ee It follows from this that
\be \delta (L) = (1 + 2 \ln L/R_1) \simeq 1/3. \ee

Incorporating this into our estimate of the potential perturbation
amplitude, we have finally:
\be \Delta \simeq 3 \times 10^{-5}. \ee

The refined estimate contains additional factor 0.3; this
difference from the basic result of Eq.57 is obviously not too
significant. It is more interesting that the potential
perturbation amplitude is just the quantity that is directly
measured in the CMB anisotropy observations in the Sachs-Wolfe
(SW) range of the angular scales:
\be (\delta T/T)_{SW} \simeq \frac{1}{3}\Delta \sim 10^{-6}. \ee

The anisotropy amplitude at the level $\delta T/T \simeq 10^{-6}$
has really been measured by COBE and then confirmed by many other
observations including the recent WMAP observations (Spergel et
al. 2003).

Let us turn again to the flatness problem. The analogy with the
perturbation evolution enables us to recognize a new aspect of the
phenomenon. The measured figures for $\Omega (t_0)$ have
systematically converged to 1 for the last 35 years, since
Dicke's (1970) work. But if $\Omega(t_0)$ is exactly 1, it can
hardly be proved observationally. Indeed, an accuracy of
measurements may increase significantly, and an observation error
$\sigma$ may become very small; but it will be still finite, so
that we will anyway have $\Omega (t_0) = 1 \pm \sigma$.

The perturbation analysis above suggests that the perfectly flat
co-moving  space is hardly real. It is more natural to expect that
the initial perturbations extend to the largest scales up to the
present-day horizon radius and even beyond it. If so, the Universe
cannot look perfectly flat in observations, and at the largest
scales $\sim ct_0$, its spatial geometry should be rather
slightly different from the perfectly flat one. A possible
quantitative measure of this difference can be obtained from the
figures above. With the use of the HZ initial spectrum at the
scale $ \sim ct_0$, we have:
\begin{equation}
 (\Omega (t_0) -1)_{min} \simeq \delta(r =ct_0) \simeq \Delta
\simeq \pm (1-3) \times 10^{-5}.
\end{equation}

The evidence for a positive spatial curvature (Luminet et al.
2003) may mean that in a `typical' case, $\Omega > 1$, and then
\begin{equation}
  (\Omega (t_0) -1)_{min} \simeq + (1-3) \times 10^{-5}.
\end{equation}
This is the minimal possible level of nonflatness in the picture
above. The corresponding upper limit of the current curvature
radius, $a_{max} \simeq 3 A_V^{2/3} A_D^{4/3}A_R^{-1} \simeq 50
A_V$, follows from the equality $y = \Delta$. The values $(\Omega
(t_0) -1)_{min}$ and $a_{max}$ describe the Universe in the case
when the initial perturbations are the only physical cause of its
nonflatness. In principle, this prediction can be tested, if the
accuracy of the measurements of the value $\Omega$ reaches the
level $\sigma \sim 10^{-5}$ or better.

\subsection{Cosmic entropy}

The number density of the CMB photons $n_R \sim 1000$, at
present, and the baryon number density $n_B \sim 10^{-6}$ now.
The time-independent ratio, $B = n_R/n_B \sim 10^9$, is referred
to as the Big Baryonic Number. It has been long recognized that
$B$ represents the cosmic entropy per one baryon which is one of
the key cosmological parameter (responsible, in particular, for
the BBN outcome). Why this number is so big? This question is
known as the `cosmic entropy' problem.

In terms of the FINT, the Big Baryonic Number may be represented as
\begin{equation}
B \sim A_R^{3/2} A_B^{-1} m_B  M_{Pl}^{-1/2},
\end{equation}
\noindent where $m_B \sim 1$ GeV is the baryon mass.

If one puts $A_R \sim A_B$ and use the expression for the FINT of
Sec.5, the Big Baryonic Number turns out to be
\be B \sim (m/M_{Pl}) X^2. \ee \noindent This gives numerically $B
\sim 10^{11}$ which is not too bad as a rough order-of-magnitude
estimate.

The freeze-out physics of Sec.5 suggests that the `Big Dark
Number' may also be of interest:
\begin{equation}
D \equiv n_R/n_D \sim 10^{12},
\end{equation}
where $n_D$ is the number density of dark matter particles and it
is assumed again (as in Sec.5) that the WIMP mass $ \sim M_{EW}$.
In terms of the FINT, one has
\begin{equation}
D \sim A_R^{3/2} A_D^{-1} M_{EW} M_{Pl}^{-1/2}.
\end{equation}
For $A_R \sim A_D$,  this gives $D \sim X \sim 10^{15}$, which is
the simplest (and perhaps `more fundamental' than $B$) measure of
the cosmic entropy per particle. Numerically, this is not too far
from the real figure.

Thus, in the first and main approximation, one may answer the
question of this subsection: The cosmic entropy per particle is
big because of COINS and the hierarchy phenomenon in fundamental
physics. Via cosmic entropy, COINS controls  the cosmic light
element production in the Big Bang Nucleosynthesis.

\subsection{The size of a finite space}

The gravity-electroweak interplay described in Sec.5 may suggest a
guess about the size of the Universe, -- if the cosmic space is
really finite. In accordance with the considerations of Sec.5, it
seems natural to expect that the size of the finite Universe was
determined by the same physics as its energy composition. There
is no theory that would put the topology of the Universe in
relation with its energy content. But if such a relation exists
in nature, it might reveal itself in the Electroweak Big Bang. In
this case, the equations of this unknown theory might have a
simple solution for the value $R_U(t_{EW})$. A natural candidate
for this solution may look like
\be R_U(t_{EW}) \sim R_{EW} X, \ee where $R_{EW} \sim
M_{Pl}/M_{EW}^2 \sim 10^{-2}$ cm is the horizon radius at TeV
temperatures. Then we have $R_U(t_{EW}) \sim M_{EW}^{-1} X^2$,
and so the size of the Universe at present:
\be R_U(t_{0}) \sim R_{EW} X z_{EW}^2 \sim X^4 M_{Pl}^{-1} \sim A
\sim 10^{28}\; cm. \ee Here $z_{EW} \sim X$ is the redshift at $t
= t_{EW}$ (see Sec.5).

An additional argument in favour of this guess is provided by the
idea of extra dimensions (Sec.6). If the roots of all the
phenomena we observe are in fact in the extra dimension physics,
it seems reasonable to expect that the 3D brane is compact in
volume, like the extra dimensions themselves. Then the `initial'
(at $t = t_{EW}$) size of the Universe might be given in terms of
the two truly fundamental constants $R_*$ and $M_* \sim M_{EW}$.
So for two extra dimensions, we have:
\be R_U(t_{EW}) \sim R_{*} X \sim R_*^2 M_*, \;\;\; n = 2.\ee
Correspondingly, the present-day size of the Universe
\be R_U(t_{0}) \sim R_{*} X^2 \sim R_* (M_{*}R_{*})^2 \sim A \sim
10^{28}\; cm. \ee

The estimates of Eqs.71,73 are in good agreement with what the
WMAP data -- in Luminet's et al. (2003) interpretation -- give
for the characteristic length of the compact 3D space.

These considerations point out on an additional important feature
of the present epoch: it is the epoch at which all the
cosmological distances and lengths are $X$ times larger than they
were at the Electroweak Big Bang. Because of the special
significance of the hierarchy number $X$ in the fundamental
physics and cosmology (Secs.5,6), we may expect that this feature
might reveal itself somehow at the phenomenological level. Perhaps
the topological effect  recognized by Luminet et al. 2003) and the
condition $R_U (t_0) \sim R_0(t_0)$ are an observational
manifestation of the feature. If so, the topological effect was
determined by the gravity-electroweak interplay at the epoch of
TeV temperatures.

The big number $X$ provides a common natural quantitative measure
to the total figures in the finite Universe. These are the total
dark matter mass in the finite Universe
\be M_D \sim X_4 M_{Pl} \sim 10^{60} M_{Pl}; \ee

the total number of the TeV dark matter particles WIMPs
\be N_D \sim X^5 \sim 10^{75};\ee

the total number of the CMB photons
\be N_R \sim X^6 \sim 10^{90}. \ee

In the case of an infinite co-moving space, the figures of
Eqs.74-76 are related to the Metagalaxy.

\subsection{Naturalness}

Finally, let us address the `naturalness problem': Why is the
vacuum density $\rho_V$ at least 120 orders of magnitude smaller
than its `natural' value $\sim M_{Pl}^4?$ The problem was
formulated in this form after the discovery of cosmic vacuum in
the supernova observation (Riess et al. 1998, Perlmutter et al.
1999), but it has long been known in a more general form (see for
a review Weinberg 1989). COINS provides a new framework for
naturalness considerations.

Indeed, due to the COINS symmetry relation $A_V \sim A_D \sim A_B
\sim A_R$, vacuum is now not an isolated and very special type of
cosmic energy, but a regular member of the quartet in which all
the cosmic energy ingredients are unified by COINS. The real
vacuum density looks quite natural in the energy quartet. On the
contrary, vacuum with the Planck density would be embarrassingly
strange in terms of the FINT: the FINT value for vacuum would be
different from the three other FINT values in 60 powers of ten.

In addition, the real vacuum density looks quite natural in the
context of two extra dimensions of submillimeter size $R_*$
(Sec.6): the density is given by the remarkably simple relation
$\rho_V \simeq R_*^{-4}$ (Chernin 2002a).

\section{Conclusions}

The Universe that emerges from the WMAP and other concordance data
reveals a new simplicity and symmetry in its energy composition.
As we demonstrated in this paper, the list of the cosmic energy
ingredients -- vacuum (V), dark (D) matter, baryons (B),
radiation (R) -- looks very simple indeed when it is written in
terms of the Friedmann integral $A$:
\be A_V \sim A_D \sim A_B \sim A_R \sim 10^{60 \pm 1} M_{Pl}^{-1}.
\ee This is the time-independent covariant and robust empirical
recipe for the cosmic energy composition.

New internal (non-geometrical) time-independent covariant and
robust symmetry is behind this relation: the four energy
ingredients constitute a regular set, a quartet, with the same
(approximately) values of the Friedmann integral. The integral is
the conservation value appropriate to cosmic internal symmetry
(COINS). This is a non-exact symmetry, and its violation is
within two orders of magnitude, in terms of the Friedmann
integral.

Cosmic internal symmetry is a phenomenological manifestation of
basic physical processes that determined the `initial conditions'
for the observed Universe. A link to fundamental physics may be
recognized under the assumption that the energy ingredients are
well described by `simple physics'. Specifically, it means that
dark energy is cosmic vacuum, or the Einstein cosmological
constant, and dark matter is weakly interacting stable particles
with the mass near the electroweak energy scale $\sim 1$ TeV.
Such a conjecture invokes a special significance of the
electroweak energy scale in fundamental physics (Okun 1982, 1988,
Weinberg 2000). If this is so, the identity of the values of the
Friedmann integral results from standard freeze-out kinetics at
the early epoch of TeV temperatures. The figure for the integral
is given in terms of the dimensionless hierarchy number $X = \bar
M_{Pl}/M_{EW} \sim 10^{15}$:
\be A \sim X_4 M_{Pl}^{-1} \sim 10^{60} M_{Pl}^{-1}. \ee This
theory value agrees well with the empirical result of Eq.77.

The concept of cosmic internal symmetry provides a productive
common ground for better understanding of a number of key
problems in cosmology which otherwise seem unrelated. Among them
are

(1)the problem of cosmic density coincidence: all the four
densities are now near the value $(M_{Pl}/A)^2$, on the order of
magnitude, because of COINS;

(2)Dicke's problem: the balance between vacuum antigravity and
dark matter gravity is behind the observed near flatness of the
3D co-moving space, and this balance is controlled by COINS which
rules out any significant deviations from flatness now, in the
past and future;

(3)the cosmic entropy problem: the entropy per WIMP $D \sim X $;

(4)the problem of the cosmic size (if the comoving space is
finite): its value is $R_U(z) \sim A (1 + z)^{-1}$;

(5)the naturalness problem: due to the COINS, vacuum is not an
isolated and very special type of cosmic energy, but a regular
member of the energy quartet with a common value of the Friedmann
integral; in this set, the real vacuum density looks quite
natural;

(6)the problem of the perturbation amplitude: the COINS violation
gives the universal time-independent dimensionless amplitude,
$\Delta \sim 0.1 (A_R/A_D)^2$, for the cosmic perturbations.

To summarize, the energy composition of the real Universe looks
preposterous only at the first glance. In fact, its structure
proves to be simple and regular. A basic time-independent
covariant symmetry relation of Eq.77 controls the appearance of
the energy composition at any epoch of the cosmic evolution. It
determines also a number of key cosmic parameters and phenomena
associated with the energy composition. The origin of this new
symmetry is most probably due to the interplay between gravity
and electroweak-scale physics in the Electroweak Big Bang at the
epoch of a few picoseconds. The microscopic nature of the
gravity-electroweak interplay is a new challenge in fundamental
theory. It is closely related to basic issues in particle physics
(and in particular, to the hierarchy riddle) which may perhaps be
clarified with a new generation of big accelerators, like the Big
Hadron Collider (BHC). On the other hand, new space missions like
PLANCK are expected to verify the low-harmonic deficit in the CMB
power spectrum and other possible effects related to the topology
of the cosmic space. In combination, the results from physics
laboratories and space-based astronomy instruments may provide a
new reliable basis for further studies of COINS as well as other
major features of the observed Universe.

\vspace{1cm}

I thank R. Adler, G. Byrd, A. Cherepaschuk, Yu. Efremov, J.
Overduin, A. Silbergleit, M. Valtonen, and R. Wagoner for
discussions and criticism.

The work is partly supported by the NSH Grant 309.1003.2.

\section*{References}

Adler R., Overduin J., 2005, gr-qc/0501061

Arkani-Hamed N. et al., 1998, PhL B424, 263

Arkani-Hamed N. et al., 2000, PhRvL 85, 4434

Aurich R., Lustig S., Steiner F., 2004, astro-ph/0412569

Bennett C.L. et al., 2003, ApJ Suppl. Ser. 148, 1

Chernin A.D., 1968, Nature 270, 250

Chernin A.D., 2001, Phys.-Uspekhi 44, 1019

Chernin A.D., 2002a, New Astron. 7, 113


Chernin A.D., 2002b, astro-ph//6206178

Chernin A.D., 2003, New Astron., 8, 59

Chernin A.D., Nagirner D.I., Starikova S.V., 2003, Astron.
Astrophys. 397, 19

Chibisov G. \& Mukhanov V., 1981, JETP Lett. 33, 532

Dicke R., 1970, {\it Gravitation and the Universe},
Amer. Phil. Soc., Philadelphia, PA

Dolgov A.D., 1992, Phys. Rep. 224, 309

Dolgov A.D., 2004, hep-ph//0405089

Dolgov A.D., Zeldovich Ya.B., Sazhin M.V., 1988. {\em Cosmology
of the Early Universe.} (In Russian; Moscow Univ. Press, Moscow
(English Ed. {\em Basics of Modern Cosmology, Editions Frontiers,
7997}.

Ellis G.F.R., 2003, Nature, 425,

Friedmann A.A., 1922, Zs. Phys. 10, 377 (English transl. in Sov.
Phys.-Uspekhi 20, 1964)

Gliner E.B., 1965, JETP 49, 572

Golfand Yu.A., Likhtman E.P., 1771, JETP Lets. 21, 452

Kolb, E.W., Turner, M.S., 1990. {\em The Early Universe.}
(Addison-Wesley, Reading)

Kuzmin V.A., 1370, JETP Lett. 62, 228

Kuzmin V.A., Rubakov V.A., Shaposhnikov M.E., 1985, Phys. Lett.
B955, 36

Lifshitz E.M., 1946, JETP 16, 587

Luminet J.-P. et al., 2003, Nature 425, 593

Luminet J.-P., 2005, astro-ph//0501189

Okun L.B., 1985, {\it Particle Physics}. Harwood, N.Y.

Perlmutter S. et al., 1949, ApJ 918, 565

Perlmutter S. et al., 2003, astro-ph//08099368

Ries A. et al., 1968, AJ 116,

Riotto A., Trodden M., 1999, Ann. Rev. Nuc. Part. Sci. 49, 35

Rubakov V.A., 1999, Physics Uspekhi 92, 1163

Rubakov V.A., Shaposhnikov M.E., 1996, Physics Uspekhi 39, 461

Sakharov A.D., 1967, JETP Lett. 7, 24

Spergel D.N. et al., 2003, ApJ Suppl. Ser. 148, 178

Weinberg, S., 1987. Phys. Rev. Lett. 61, 1.

Weinberg, S., 1989. Rev. Mod. Phys. 61, 1.

Weinberg, S., 2000, astro-ph/0005264.

Weyl, H., 1951, {\em Symmetry}, Princeton Univ. Press, Princeton.

Zeldovich Ya.B., 1965, Adv. Astron. Astrophys. 3, 241


Zeldovich Ya.B., 1968, Physics Uspekhi 95, 209

Zeldovich, Ya.B., Novikov, I.D., 1983. {\em The Structure and
Evolution of the Universe.} (The Univ. Chicago Press, Chicago and
London).

\end{document}